\documentstyle[11pt,newpasp,twoside,epsfig]{article}

\def \lesssim{\mathrel{<\kern-1.0em\lower0.9ex\hbox{$\sim$}}}
\def \gtrsim{\mathrel{>\kern-1.0em\lower0.9ex\hbox{$\sim$}}}

\begin{document}


\title{Formation of Compact Binaries in Globular Clusters}
\author{Saul Rappaport, Eric Pfahl, Frederic Rasio}
\affil{Department of Physics, Massachusetts Institute of Technology,
Cambridge, MA 02139}

\author{Philipp Podsiadlowski}
\affil{Nuclear and Astrophysics Laboratory, Oxford University, Oxford,
OX1 3RH, England, UK}


\begin{abstract}
Globular clusters are overabundant per unit mass in neutron-star binaries,
including low-mass X-ray binaries and radio pulsars.
We report here on two complementary
population synthesis studies which relate directly to the formation and
evolution of neutron star binaries in globular clusters.  In the first we
consider an early population of massive stars, including those in binary
systems, which produce the original population of neutron stars.  We
compute the probability of retention of these neutron stars, and
quantitatively confirm the idea that the retention fraction for neutron
stars born in binary systems is greatly enhanced over those born in
isolated stars.  However, with the currently fashionable natal kick
velocities given to neutron stars, the retention fraction may well be {\rm
in}sufficient to explain the current population of neutron star binaries.
In the second study, we follow a large population of primordial binaries
and neutron stars throughout the lifetime of a globular cluster whose
properties may be similar to 47 Tuc.  We directly compute all 3-body
interactions among binary systems, neutron stars, and isolated field stars
throughout the history of the cluster.  The evolution of certain types of
neutron star binaries is followed up to the current epoch.  The numbers of
close, recycled, binary radio pulsars are evaluated and compared with the
results of radio observations.   We find that for an initial population of
some $10^6$ primordial binaries and $\sim 10^4$ neutron stars, we can
plausibly account for the large numbers of binary radio pulsars that are
inferred to exist in clusters such as 47 Tuc.
\end{abstract}


\section{Introduction}


Globular clusters are known to contain a large variety of highly
interesting systems containing neutron stars (NSs) and white dwarfs (WDs).
These include binary millisecond pulsars, low-mass X-ray binaries, cooling
WDs in binaries, CVs, and the enigmatic low-luminosity X-ray sources.
Clusters are overabundant in these objects, per unit mass, due to the fact
that numerous dynamical formation channels exist in clusters that are not
available to stars born in the galactic disk.  We describe here a
comprehensive population synthesis study of the production and retention of
compact objects in clusters and the binary systems containing these
objects.  Our study differs from others in that we follow an entire
population of primordial binaries and NSs through the lifetime of
their host cluster.

The study divides itself naturally into two parts.  In the first part, we
consider an early cluster population of massive stars, including those in
binary systems, which produce the original population of NSs.  We
apply Monte Carlo methods, binary stellar evolution theory, and natal kicks
to the NSs to calculate the retention probability for these
NSs.  We compute the NS retention fraction as a
function of the rms kick speed and the central escape speed of the globular
cluster.

In the second study, we follow a large population of primordial binaries
and NSs throughout the lifetime of a globular cluster whose
properties may be similar to 47 Tuc.  Each of the NSs and primordial
binaries is followed as it sinks toward the cluster
core region as a result of dynamical friction.   In or near the core, the 
binaries are allowed to undergo
randomly chosen (according to the appropriate cross sections) collisions
with field stars and NSs.  We directly compute
all relevant 3-body interactions among these systems throughout the history
of the cluster.  All of the binaries and $\sim 10^4$ NSs are followed, in
parallel, as a function of time to the present epoch.  The evolution of
certain types of NS binaries are followed in detail up to the
current epoch.  The numbers of close, recycled, binary radio pulsars are
evaluated and compared with the findings of radio observations.   We
conclude that for an initial population of some $10^6$ primordial binaries
and $\sim 10^4$ neutron stars, we can plausibly account for the large
numbers of binary radio pulsars that are inferred to exist in clusters such
as 47 Tuc.  Companions to the NSs turn out to be white dwarfs,
other NSs, and normal stars with lower than the turnoff mass.   Other
interesting binaries include combinations of He and CO WDs.


\section{Retention of Globular Cluster Neutron Stars}\label{B}

A major question to be answered in the study of NS binaries in
globular clusters is, ``How are the NSs that are born in the
cluster retained by the cluster?"  The problem is that globular clusters
have typical central escape speeds of a few 10's of km s$^{-1}$, and neutron
stars are widely thought to be born with substantial ``kick" velocities.  A
three-dimensional speed distribution for NSs in the galactic disk
has been inferred from proper motion studies of about 100 radio pulsars
(e.g., Lyne \& Lorimer 1994).
A number of authors (e.g., Hansen \& Phinney 1997) have taken
the natal kick speed distribution to be represented by a Maxwellian of the
form $p(v) \propto v^2 e^{-v^2/2\sigma^2}$,
where $\sigma\simeq$ 190 km s$^{-1}$.  A simple integration of this
distribution from $v = 0$ to the typical escape speed of a globular
cluster quickly demonstrates that only a very small fraction of neutron
stars receiving such a kick could remain bound in the cluster ($\sim 0.1 - 1\%$
for a reasonable range of escape speeds).  This is the essence of the
NS ``retention" problem.

It has been suggested by a number of authors, and demonstrated
quantitatively (e.g., Drukier 1996; Davies \& Hansen 1998), that if
NSs are born in massive binaries, then there is a significant
probability that the NS would remain in the binary, and
that the recoil of the binary could be sufficiently small to allow it to
remain bound in the cluster.   In a study of a grid of models,
Davies \& Hansen (1998) found that somewhere between $\sim4\%$ and
$\sim25\%$ of the NSs formed in massive binaries could be retained
in a typical cluster, depending on the initial binary parameters. While
these studies provided a useful verification of the potential role
of massive binaries in retaining NSs in clusters, they did not
involve a population study to determine a realistic net neutron
star retention fraction. This is crucial, since the retention fractions found
in the Davies \& Hansen (1998) study span the range from providing the
requisite number of NSs to leaving the problem essentially
unsolved.

We utilize Monte Carlo techniques to choose the initial parameters for some
large number (e.g., $10^6$) of primordial binaries (see, e.g., Rappaport,
Di Stefano, \& Smith 1994; Soker \& Rappaport 2000).  The primary mass is
chosen from an assumed initial mass function (IMF) (see, e.g., Miller \& Scalo
1979; Kroupa, Tout, \& Gilmore 1993).  For this
part of the study, we consider only primaries sufficiently massive (i.e.,
$\gtrsim8\,M_\odot$) to form NSs.  We adopt a distribution of
mass ratios for primordial binaries, $p(q)$, which is approximately flat
(see, e.g., Duquennoy \& Mayor 1991).  The initial orbital period,
$P_{orb}$, is chosen from a function which is constant in log $P_{orb}$
(see, e.g., Duquennoy \& Mayor 1991; Abt \& Levy
1985).  The orbital eccentricity is chosen from a uniform distribution.
At the onset of mass transfer we simply assume that the orbit circularizes
and that orbital angular momentum is conserved in the process.

Once the parameters of a primordial binary have been set, the subsequent
evolution of the primary and the effects of mass transfer and loss are
followed according to a set of prescriptions (see, e.g., Podsiadlowski,
Joss, \& Hsu 1992).   Wherever possible, these prescriptions are derived
from stellar evolution calculations with a Henyey-type code (Kippenhahn,
Weigert, \& Hofmeister 1967).  Once the
primary evolves, there are a number of possible outcomes, depending on
whether mass transfer takes place while the primary is still on or near the
main sequence (Case A), after the core is hydrogen exhausted (Case B), or
after the core is helium exhausted (Case C); there is also the possibility
that the orbit is sufficiently wide that no mass transfer takes place.  For
the case where mass transfer from the primary to the secondary takes place,
the transfer may be either stable or dynamically unstable, depending on the
mass ratio of the two stars and the evolutionary state of the primary.

For stable mass transfer a prescription is needed for the fraction,
$(1-\beta)$, of transferred mass that is ejected from the binary system and
the specific angular momentum, $\alpha$, that is carried away with the
ejected matter (see Podsiadlowski, Joss, \& Hsu 1992).
The value of $\beta$ is rather uncertain and may essentially be taken to
be a free parameter of the problem.  At present we take a fixed value of
$\alpha$ = 1.5, which is close to the value expected for matter ejected
from the L2 point.  For $\beta$ we somewhat arbitrarily take a fixed value
of 0.7.

If the mass transfer is dynamically unstable, we assume that a
common envelope (CE) phase is initiated and that the secondary spirals in
toward the core of the primary as a result of drag forces.  We adopt the
standard approach of assuming that some fraction, $\alpha_{CE}\sim1$, of the
initial orbital binding energy is deposited into the CE as frictional
luminosity (see, e.g., Meyer \& Meyer-Hofmeister 1979; Sandquist, Taam, \&
Burkert 2000). For the envelope binding energy we use the recent
calculations of Dewi \& Tauris (2000) and Han, Podsiadlowski \& Eggleton
(1995; who also consider the ionization energy in the envelope).
If sufficient energy is available so as to unbind the
envelope, what remains is a compact binary consisting of the secondary,
which we assume is unaltered in the spiral-in process, and the helium-rich
core of the primary.  If the CE is not ejected, then drag forces will
perpetuate the spiral-in until the two stars merge.

At the end of this first part of the evolution the result may be a stellar
merger, a He star in orbit with a companion, or a wide binary whose
evolution is essentially that of two single stars.  Whichever the case, the
primary evolves to core collapse, followed by a supernova explosion.
At this point we choose a natal kick speed from some distribution
(e.g., a Maxwellian) and a random direction for the kick.
The post-supernova orbit is computed, and if the NS escapes the
binary, its speed relative to that of the pre-supernova binary is computed,
while if the NS remains bound in the binary, the recoil of the
binary is computed.  In either case it can be established
whether the NS (either single or in a binary) is retained in the
cluster.  At the present time we simply use a fixed escape speed from the
cluster, i.e., we do not consider the location of the binary within the cluster
or the velocity of the binary prior to the supernova.

After this process is repeated for $\sim10^{5}$ primordial binaries, we
simply count the fraction of NSs retained by the cluster, whether
single or in binary systems.  As the binaries containing NSs
continue to evolve, it is likely that the NS will ultimately be
engulfed in the envelope of the companion once mass transfer commences
(i.e., following a brief interval as a high-mass or intermediate-mass X-ray
binary).  In many cases, especially for the shorter orbital period systems
where the companion is not too evolved when mass transfer occurs, this will
lead to a complete spiral in and the possible formation of a Thorne-\.Zytkow
object (Thorne \& \.Zytkow 1977).  The final fate of these objects is
the subject of considerable debate (Podsiadlowski, Cannon, \& Rees 1995).
The envelope of the Thorne-\.Zytkow object may ultimately be dispersed,
leaving a NS, or the NS may undergo hypercritical accretion
and collapse into a black hole (Chevalier 1993; Bethe \& Brown 1998).  Since
a second common-envelope phase may constitute a significant path for retaining 
NSs, the calculations we are performing could shed important light on the answer to
this question.

\begin{figure}
\noindent
\begin{minipage}[c]{0.6\linewidth}
\centering\epsfig{file=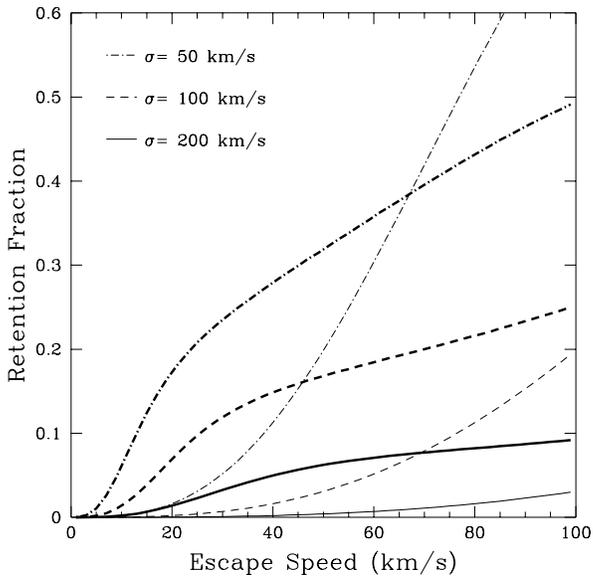,width=\linewidth}
\end{minipage}
\hfill
\begin{minipage}[c]{0.4\linewidth}
\caption{Retention of NSs in globular clusters.  Heavy curves
are for NSs born in binary systems; light curves represent NSs
formed from single stars.  The rms kick speed used in the 
Maxwellian distribution is denoted by $\sigma$.}
\label{fig:retention}
\end{minipage}
\end{figure}

Some preliminary results from our retention study are given
in Figure \ref{fig:retention}.  The plot shows the fraction of NSs
retained in a cluster as a function of the escape speed from the cluster.
The thick curves are for systems born in binary systems while the thin
curves are for NSs born from single stars.  The sequence of three
line styles labeled 50, 100, and 200 km s$^{-1}$ corresponds to the value of
$\sigma$ used in the Maxwellian kick speed distribution.  Note that
for NSs formed from single stars and kick speeds characterized by $\sigma =
200$ km s$^{-1}$, the retention fraction is very small (i.e., $\lesssim2\%$)
which is insufficient to explain the numbers of NSs that are
inferred to be in globular clusters (Kulkarni, Narayan, \& Romani 1990;
Davies \& Hansen 1998; Camilo et al. 2000). The formation of NSs in binary
systems, with $\sigma = 200$ km s$^{-1}$, significantly increases the
retention fraction to $\lesssim8\%$, but even this probably falls short of
the required efficiency.

Only when the value of the mean Maxwellian kick speed is reduced by a
factor of $>2$ below the conventionally accepted value does the
retention efficiency via binaries increase to the point where a healthy
population of NSs can be retained. One immediate, albeit tentative,
conclusion is that there is an apparent need for a significant low-speed
component to the kick distribution.  Thus, it seems difficult
to reconcile the apparent numbers of NSs in clusters, with the
conventionally adopted natal kick speed distribution for NSs (Lyne \&
Lorimer 1994; Hansen \& Phinney 1997; Cordes \& Chernoff 1998).



\section{Formation of Binaries With Neutron Stars}


\label{C}

As mentioned above, the overabundance of NS binaries in globular
clusters per unit mass, compared to that in the galactic plane, has led to
the general conclusion that these objects are not formed directly from
primordial binaries, but rather are the result of NS captures of
a companion star.  There are a number of scenarios that have been discussed
over the years in which NSs can capture a companion star, in
particular, low- and intermediate-mass stars which are the preferred
companions for forming sources referred to as ``LMXBs" and recycled
pulsars. These include: (i) tidal capture of single field stars (Fabian,
Pringle, \& Rees 1975) and (ii) exchange collisions between primordial
binaries and NSs (Hut, Murphy, \& Verbunt 1991; Sigurdsson \& Phinney 1993).

Tidal capture of a main-sequence or giant star by a NS has been
invoked to explain the formation of close binaries in globular clusters
(see, e.g., Di Stefano \& Rappaport 1992).  However,
there are a number of problems with this scenario including possible
catastrophic dynamical effects encountered during the initial capture and
subsequent few orbits (Rasio \& Shapiro 1991), as well as potentially
catastrophic tidal heating during the subsequent circularization process
(Ray, Kembhavi, \& Antia 1987; Podsiadlowski 1996).  However, the tidal heating
may not present such a problem for the capture of more massive stars
early in the cluster's history  (Podsiadlowski 2000).

In the remainder of this paper we concentrate on the formation of
NS binaries via NS-binary interactions.



\vspace{3mm}
\noindent{\bf Binary-Neutron Star Encounters} 
Our population synthesis code for NS
binaries in globular clusters includes a 3-body
integrator with which we explicitly follow all relevant binary-single
star encounters over the history of the globular cluster.   From the total cross
section for a ``strong'' 3-body encounter we
choose via Monte Carlo methods whether there is, or is not, a scattering
event for each binary in each time step.  
(A ``strong'' encounter is one in which the distance of closest
approach is equal to $\lesssim3$ times the apastron separation of the binary.)
If there is a strong encounter, then we choose
whether the incident single star is a NS or a field star.  If the latter, the
mass and type of object (either a normal star or a WD) are chosen from
the stellar mass function at that epoch. Finally,
the incident configuration and impact parameter for the scattering event
are chosen.

A typical population synthesis run, which follows $10^6$ binaries and $10^4$
NSs, takes 2 days on a Sun Ultra 5 workstation; this includes
following $\sim2 \times 10^6$ scattering events.

During some binary-single star encounters, especially for those incident
binaries with shorter orbital periods (e.g., $\lesssim$ a few days), and for
those which develop into long-lasting resonant scattering events, there is
an enhanced probability for two of the stars to come quite close together.
In this case, ``close" is defined as leading to a significant tidal
interaction, or even mass transfer.  At the present time, we merely keep
track of these occurrences and eliminate systems where a NS would
come close enough to one of the other stars to drive mass transfer. From our 
preliminary studies we have found that for the great {\it majority}
of the scattering events we follow, there are no tidal or mass-transfer
encounters between any of the stars; however, a non-negligible fraction of
scattering events do show close interactions, and these should provide some
interesting cases to study in the future.  The reason for the relatively small
number of close encounters lies in the fact that most of the 3-body
collisions we follow involve systems with orbital periods $\gtrsim10$ days,
with some up
to 100's of days.

Due to the possibility of a large fraction of primordial binaries in
globular clusters (Hut et al. 1992), binary-binary interactions
may also be important in the production of NS-binaries and in
the partial depletion of the binary population.  We plan to incorporate
binary-binary interactions in the future.  For the present study, however,
all of our results are based on only binary-single star encounters.


\subsection{Stellar Evolution Within Systems Containing Neutron
Stars}\label{G}


Any given primordial binary may undergo a complicated series of exchange
interactions (possibly also an ionizing interaction which terminates that
particular binary), one or more of which may leave a NS in the
binary.  If, at some point in the history of that binary, before the
current epoch, the companion to the NS evolves to the point of
filling its critical potential lobe at periastron passage, the subsequent
mass transfer evolution must be followed in order to ascertain whether an
interesting binary, e.g., a LMXB or recycled radio pulsar, is formed.

The onset of mass transfer divides into two distinct branches: dynamically
unstable and dynamically stable.  In the former case a CE
ensues and the NS spirals in toward the core of the donor star.
The net result is likely to be the formation of a very close binary pair
consisting of a NS and WD with an orbital period of a
fraction of a day.  [This assumes, of course, that the NS does
not accrete sufficient material to collapse into a black hole during the
spiral-in process (Chevalier 1993; Bethe \& Brown 1998).]
If the post-CE orbital period is sufficiently short (typically $<6$ hours
for a WD-NS pair) gravitational radiation losses may bring the binary to a
semi-detached state, with the WD filling its Roche lobe, by the
current epoch.  The subsequent evolution will be dynamically
unstable if the mass of the WD exceeds $\sim0.4~M_\odot$.
Otherwise the mass transfer will be quite rapid, but nonetheless stable.
We discuss below this type of short-lived X-ray source as possible
progenitors of the very short orbital period binary radio pulsars discovered
in 47 Tuc.

The other distinct possibility is that the initial mass transfer from
the normal donor star to the NS will be {\it stable}.  At the present time
we do not attempt to follow the evolution of binary systems leading to
stable transfer.  We merely store the properties of such binaries at the onset
of mass transfer.

To determine mass transfer stability from a main-sequence, subgiant, or giant
star onto a NS, we combine a number of results to
produce an analytic expression for the ``adiabatic" stellar index,
$\xi_{ad}$, of such a star, i.e., the logarithmic derivative of radius with
respect to mass of the donor star in the absence of any heat input to the
star.  In particular, we used results from Hjellming (1990; see also Kalogera
\& Webbink 1996), as well as from our own stellar models with constant
mass loss rates, as inputs to deriving such an expression.  In
our formulation $\xi_{ad}$ is a function of the instantaneous mass of the
donor, its original mass, and its evolutionary state (e.g., core mass or
composition). We also ensured that in certain well-known limiting cases, e.g.,
for a fully convective star, $\xi_{ad}$ goes to the correct limit  (e.g.,
$\xi_{ad}= -1/3$).



\vspace{3mm}
\noindent {\bf Evolution of the Common-Envelope Products}
If the result of the initial mass transfer onto the NS is
unstable, leading to a CE phase and the production of a WD-NS binary
(as discussed above), the subsequent evolution of
such a system can be quite interesting and may well explain some of the
LMXBs and short period recycled radio pulsars in globular clusters
(see the evolutionary schematic in Fig. \ref{fig:schematic}).  For a
detached WD-NS binary the orbital decay timescale via
the emission of gravitational radiation is given by
$\tau_{GR}= 3 \times 10^{7} (P_{orb}/{\rm hr})^{8/3}(\mu/M_\odot)^{-1}
(M_{tot}/M_\odot)^{-2/3}$ yr, where $\mu$ and $M_{tot}$ are the system
reduced mass and total mass, respectively.
So, for typical WD masses of $0.2 - 0.4\,M_\odot$, systems with
$P_{orb} \lesssim6$ hrs have a reasonable probability of becoming
semi-detached (i.e., the WD fills its Roche lobe) between the
formation of the binary and the current epoch.  Once the WD starts
to transfer matter to the NS it will be driven at very high rates
by gravitational radiation losses (far in excess of the Eddington limit),
and therefore mass will be ejected from the system (i.e., $\beta\simeq
0$).  The condition for stability is that the quantity
$D=5/6+\xi_{ad}/2-(q+3)/[3q(1+q)] > 0$ (Rappaport, Verbunt, \& Joss
1983), where $\xi_{ad}$ is the adiabatic stellar index of the WD
(see the above discussion), and $q \equiv M_{ns}/M_{wd}$.
In turn, $\xi_{ad}\simeq(-1/3)[1+m^{4/3}]/[1-m^{4/3}]$ where $m \equiv M_{wd}/
1.4~M_\odot$ (see, e.g., Rappaport et al. 1987).  If we combine these two
expressions, we find that essentially all He WDs will transfer
mass to a NS companion stably, while CO WDs will not.

\begin{figure}
\noindent
\centering\epsfig{file=rappaport_binev.epsi,height=\linewidth,angle=-90}
\caption{Schematic evolution of a highly compactWD-NS binary.}
\label{fig:schematic}
\end{figure}

For a WD-NS binary undergoing dynamically unstable mass transfer, the
result will be a massive disk formed around the NS.  The outcome
of this is greatly uncertain, but it may be a spun up NS or black
hole.  We simply record these events but do not attempt to compute what the
end product might be.

\begin{figure}
\noindent
\begin{minipage}[c]{0.6\linewidth}
\centering\epsfig{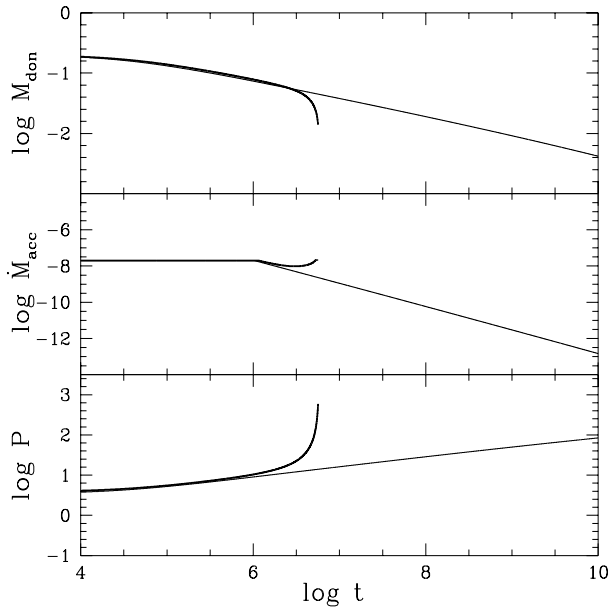}
\end{minipage}
\hfill
\begin{minipage}[c]{0.4\linewidth}
\caption{White dwarf-neutron star binary evolution.  Solid curves are for the
case of no tidal heating.  Heavy curves include tidal heating; asynchronous
rotation of the WD with the orbit is maintained artificially.
The units for the three panels are $M_{\odot}$, $M_{\odot}\,{\rm yr}^{-1}$, and 
minutes, respectively.}
\label{fig:WD-NS}
\end{minipage}
\end{figure}

For WD-NS systems with stable mass transfer, the binary evolution is
straightforward to follow for the case where the WD remains
degenerate.  A sample evolution plot for a WD-NS binary is shown in Fig.
\ref{fig:WD-NS} (thin solid curves). The initial mass of the WD is
$0.2\,M_{\odot}$, but the evolution of all stable WD-NS binaries join
together once the respective masses of the WDs have been reduced to a
common value.  Note that $M_{wd}$ and ${\dot M}$ decrease rapidly at the
start of mass transfer, then slow considerably as the evolution progresses.
During the same time, the orbital period grows due to the fact that
the donor star is less massive than the accretor, even though angular
momentum is being extracted from the binary via gravitational radiation.
The overall evolution time for the binary to expand back to a period of
$\sim2$ hrs is of the order of a Hubble time.

This basic evolutionary scenario appears promising to explain 5 of the
binary radio pulsars in 47 Tuc with short orbital periods (Camilo et al. 2000)
and very low-mass companions ($\sim0.02\,M_\odot$), as well as 3 LMXBs in
other globular clusters; however, the values of $P_{orb}$ for some of these
recycled pulsars range up to $\sim6$ hr, which cannot be reached with mass
transfer driven by gravitational radiation alone.  We have therefore also
considered the effects of tidal heating as an added source of internal
energy to keep the WD ``bloated" beyond its completely degenerate radius
(Applegate \& Shaham 1994).  Inclusion of tidal heating is somewhat
speculative and requires the assumption of an (unknown) mechanism for
keeping the WD rotation asynchronous with the orbit, as well as a
synchronization timescale (a measure of viscous damping) which
is essentially a free parameter.  We have used a simple polytrope to
represent the WD in carrying out the tidal evolution calculations,
and no surface cooling has been taken into account. 
For the purposes of this calculation, we have adopted a fixed {\it a}synchronization
factor of 50\% and a synchronization timescale of $6 \times 10^4$ yr
(see Rasio, Pfahl, \& Rappaport 2000). 

The results of the tidal evolution calculations are shown in
Figure \ref{fig:WD-NS} as the set of heavy curves.  The tidal heating
results in a rapidly accelerated evolution, and a larger orbital period is
reached than can be attained with gravitational radiation alone.  Not only
are the longer periods important (to match the observations), but the X-ray
phase is terminated more quickly.  This bears on the question of how many LMXBs
with short orbital periods should be found in clusters at the current
epoch.  At present, only 4U1820-303 ($P_{orb}$=11 min; Stella
et al. 1987), X1850-087 ($P_{orb}$=21 min; Homer et al.
1996), and X1832-330 ($P_{orb}$=44 min; Deutsch, Margon, \& Anderson 2000 
and references therein) may correspond to this type of system.  This modest number 
is, in fact, more compatible with a short lived X-ray phase, than one without tidal heating.

We note that there is an alternative channel for forming compact
NS binaries that does {\it not} involve a CE phase.  From our binary
evolution calculations, we find that if, at the start of mass transfer, the
donor star is near the main sequence ($P_{orb}\lesssim1~{\rm day}$) and
its mass is in the range of $\sim 1-3~M_\odot$, then the mass transfer
proceeds stably on a thermal timescale and the orbital period tends
to shrink as magnetic braking removes angular momentum
from the orbit, yielding periods as short as $\sim5$ min.  However,
the relevance of this scenario for the formation of compact binaries in
globular clusters is not yet clear since, at present, our simulations
indicate that most binaries containing a NS commence mass transfer
when $P_{orb} \gtrsim 10$ days.



\vspace{3mm}
\noindent {\bf Systems Undergoing Stable Mass Transfer}
For captured field stars undergoing {\it stable} mass transfer onto the neutron
star, we would like to be able to follow this portion of the binary
evolution in detail so that we can (i) classify the system at various
epochs according to its properties as an X-ray source, and (ii) learn what
sort of remnant recycled binary radio pulsar will remain at the current
epoch.  However, this would require following the evolution of a large number
of systems with a Henyey-type stellar evolution code, which is very time
consuming.  This is clearly something that should be done in future
studies.

\begin{figure}
\begin{minipage}[c]{0.6\linewidth}
\centering\epsfig{file=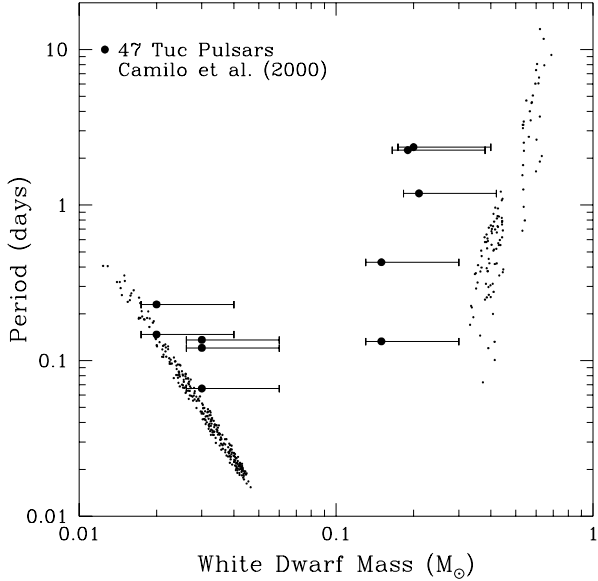,width=\linewidth}
\end{minipage}
\hfill
\begin{minipage}[c]{0.4\linewidth}
\caption{Orbital period vs. white dwarf mass for the end products
of systems involving a low-mass WD and a NS.}
\label{fig:newp-mwd}
\end{minipage}
\end{figure}


\subsection {Other Considerations for the Binary Population Synthesis Study}



\noindent {\bf Time-Dependent Aspects of the Overall Cluster Evolution}
At least 80\% of globular clusters in our Galaxy have well-resolved cores
and are well-fitted by standard King models (e.g., Djorgovski 1993). These
clusters are thought to be supported against core collapse by the heating
generated in their cores through dynamical interactions involving hard
primordial binaries (Gao et al.\ 1991; Hut et al.\ 1992; McMillan
\& Hut 1994; Rasio 2000). During this 
``binary-burning" phase of cluster evolution, the core
parameters remain nearly constant in time. In particular, the core radius
and central density do not vary by more than a factor of $\sim 2$.
This justifies our assumption of an approximately steady-state cluster
density profile.



\vspace{3mm}
\noindent {\bf Sinking Times of Binaries and Neutron Stars}
At present, we assume that binaries and NSs undergo mass segregation and enter
the cluster core in a time $t_s$, distributed according to
$p(t_s)=(1/t_{sc}){\rm exp}(-t_s/t_{sc})$, where the characteristic time
$t_{sc}\simeq10(m/m_f)t_{rh}$ for objects of mass $m$ drifting
through field stars of average mass $m_f$ (see, e.g., Fregeau, Joshi, \&
Rasio 2000).  For the present study, we have simply adopted a fixed value
for $t_{rh}=10^9$ yr.


\vspace{3mm}
\noindent {\bf Temporal Evolution of the Field Star Population} 
We assume that the
background of single objects (normal stars and WDs) in the globular cluster
core is completely confined to the core for the entire lifetime of the cluster.
Thus, a star more massive than the cluster turnoff mass will shed its envelope
to become a WD, and we assume that the WD remains in the core.  Based on this
{\em confinement} assumption we have derived an essentially analytic
formula for the time-dependent mass function of field stars in the core,
given the {\em initial} mass function of the stellar population.  This
includes a developing population of degenerate remnants.


\subsection{Results From the Population Synthesis Study}\label{M}

\noindent
Here we show some examples of the types of results that can be obtained
with the approach we have adopted.  The cluster parameters used to
produce the following results were:
$10^6$ primordial binaries; $10^4$ neutron stars, and a constant core
density of $10^5$ stars pc$^{-3}$ (i.e., a possible representation of 47
Tuc).

\begin{figure}
\noindent
\begin{minipage}[t]{0.44\linewidth}
\centering\epsfig{file=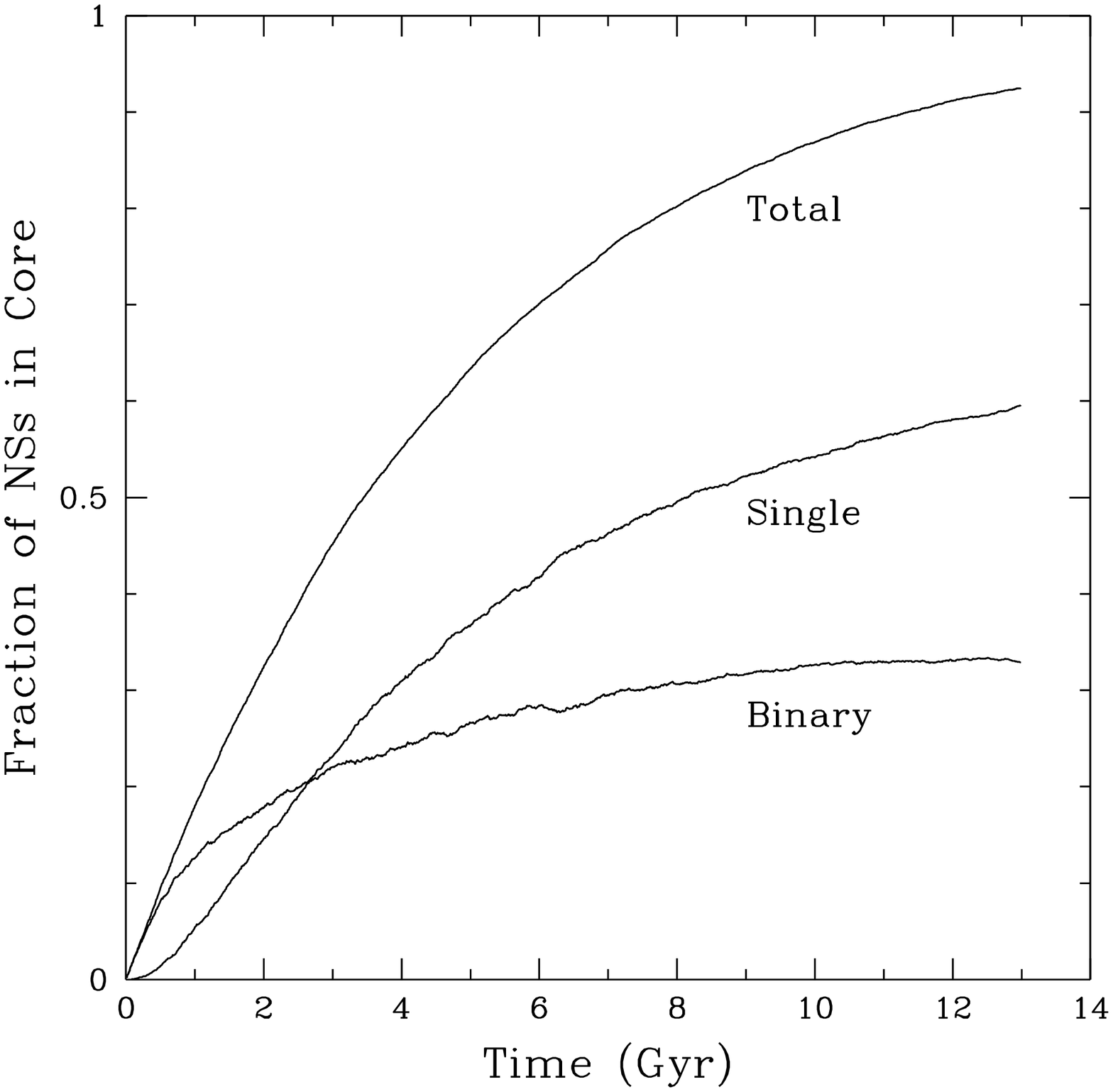,width=\linewidth}
\caption{Fraction of the $10^4$ NSs, both single and in binaries,
in the core of the simulated globular cluster.}
\label{fig:numincore}
\end{minipage}
\hfill
\begin{minipage}[t]{0.44\linewidth}
\centering\epsfig{file=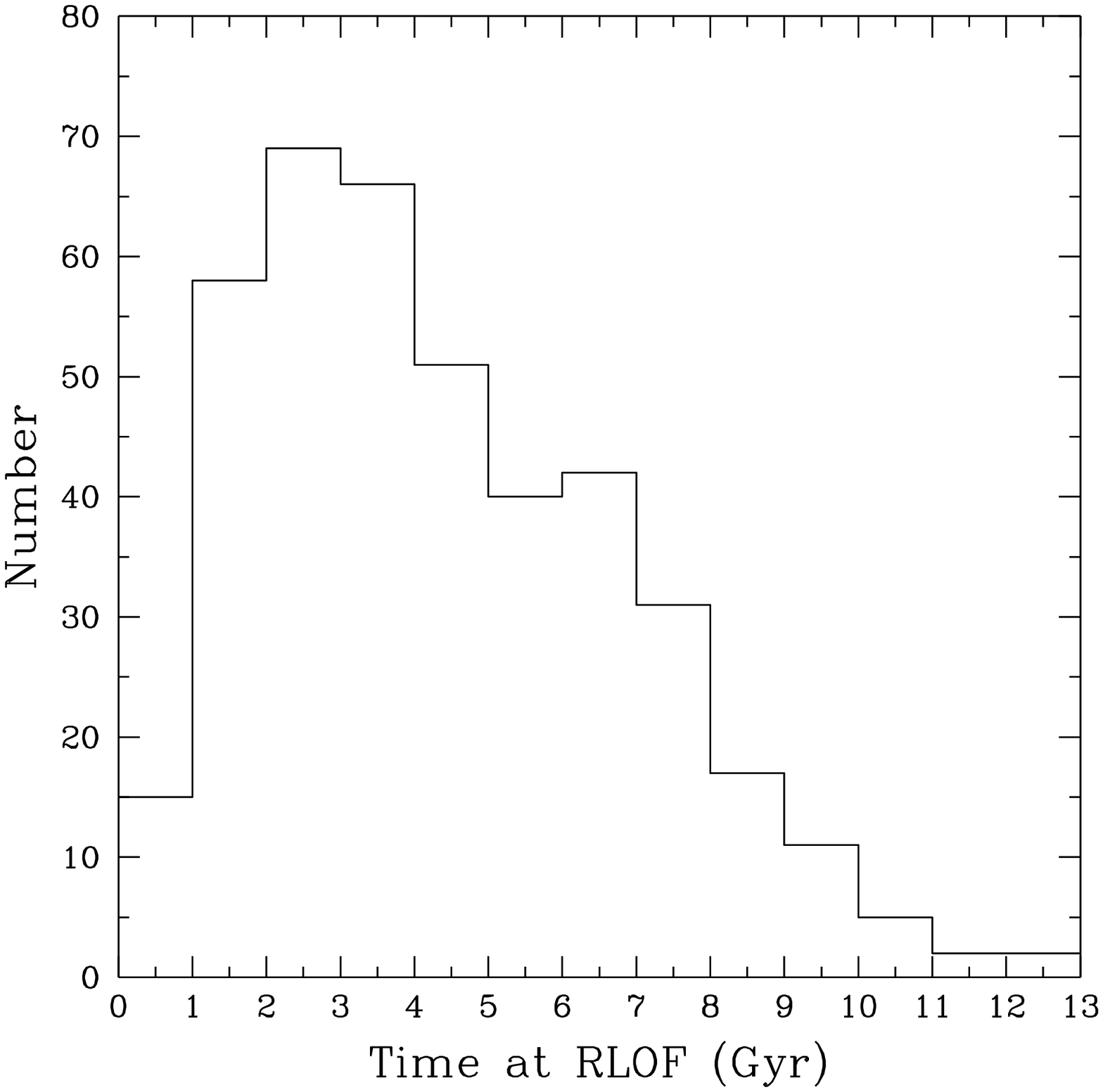,width=1.03\linewidth}
\caption{Time of first RLOF for systems becoming
white dwarf-neutron star binaries.}
\label{fig:RLOF}
\end{minipage}
\end{figure}

In Figure \ref{fig:newp-mwd} we show the simulated population of NS-WD
binaries at the current epoch in the $P_{orb}-M_{wd}$ plane (see Rasio,
Pfahl, \& Rappaport 2000).  Note the ``track" of systems to the lower
left which have $20~{\rm min}\lesssim P_{orb}
\lesssim~6$ hr and WD companion masses in the range of 0.01--0.04
$M_\odot$.  These have all been spun up by the accretion process
to rotation periods shorter than 5 msec. This track lies
suggestively near a group of 5 binary millisecond pulsars discovered in 47 Tuc
(Camillo et al. 2000).   The other prominent group of simulated systems
lies toward somewhat longer orbital periods ($2~{\rm hr}-10$ days) and
$M_{wd} \sim 0.4~M_\odot$.  While these systems do lie in a part of
parameter space near to 5 other binary msec pulsars found in 47 Tuc, the
computed binaries have systematically larger companion masses and,
moreover, should not have undergone any significant amount of accretion to
spin up the pulsar (these are systems where the WD-NS binary never decayed
sufficiently for mass transfer to commence).

The simulated systems shown in Fig. \ref{fig:newp-mwd} represent only a portion
of the types of interesting systems formed in the population synthesis, i.e.,
those that result from unstable mass transfer from the normal companion to
the NS leading to a CE phase.  Not shown are systems of
double WDs and double NSs, and importantly, none of the systems where the
mass transfer from the normal star to the NS is stable.  In future studies
we plan to follow all of the systems undergoing stable mass transfer either
with a Henyey-type code or with a large grid of pre-prepared binary
evolution models.

The code also produces a wealth of supplementary information,
some of which we illustrate here.  From the distribution of the number
of scattering interactions that a binary system undergoes before
either one of the stars commences Roche-lobe overflow (RLOF)
or the current epoch is reached, we find that the average number of
interactions
is $\sim25$, while some binaries undergo more than 100 interactions.
Fig. \ref{fig:numincore} shows the number of NSs (labeled
``total'') as a function of time in the core of the cluster; the curves labeled
``single'' and ``binary'' are for isolated NSs vs. those in
binaries.  In Fig. \ref{fig:RLOF} we show a histogram of the number of
binaries (containing a NS) which commence mass transfer as a
function of time.  Note that this distribution peaks at $\sim3$ Gyr and is
rather low by the current epoch.  Therefore, most of the incipient mass
transfer systems were of intermediate mass and occurred relatively early in
the history of the cluster (see also Davies \& Hansen 1998). This type
of result bears directly on the issue of the``birthrate problem" (Kulkarni,
Narayan, \& Romani 1990) since the progenitors of many of
the recycled pulsars may have been intermediate-mass X-ray binaries (IMXBs)
rather than LMXBs (Davies \& Hansen 1998), were born long ago in the history
of the cluster, and the X-ray phase of such systems may no longer be observed.

\noindent
\begin{figure}
\begin{minipage}[c]{0.6\linewidth}
\centering\epsfig{file=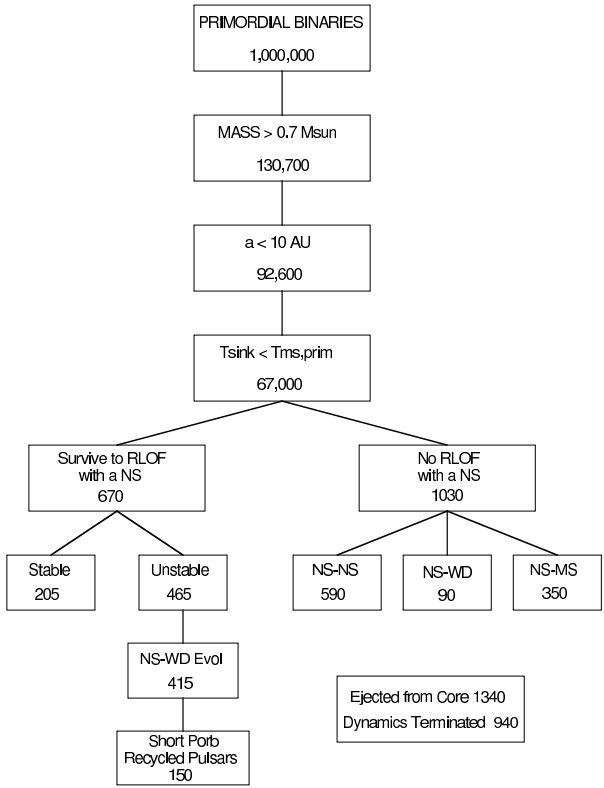,width=\linewidth}
\end{minipage}
\hfill
\begin{minipage}[b]{0.4\linewidth}
\caption{Numbers of surviving systems in the population synthesis
study that contain neutron stars.}
\label{fig:flow}
\end{minipage}
\end{figure}

Finally, we show in Figure  \ref{fig:flow} a ``flow" diagram which illustrates
what happens to the $10^6$ primordial binaries and $10^4$ NSs that
we started with in the particular run which generated the figures shown
above.  After the cuts requiring a minimum mass for the primary
($0.7~M_\odot$), a maximum semimajor axis of 10 AU (such systems would
be quickly ionized), and no RLOF before the primordial binary sinks to the
core, there are $6.7\times10^{4}$ binaries remaining.  Roughly
2000 binaries acquire a NS via exchange interactions;
of these about 1/3 ultimately undergo RLOF from an evolving donor star.
About 2/3 of the RLOF cases are unstable and produce WD-NS binaries,
while the remainder undergo stable mass
transfer.  Of the systems which end up without RLOF
onto the NS, there are substantial numbers with companion
WDs ($\sim100$), NSs ($\sim600$; these are generally in
wide orbits and are not recycled), and low-mass, unevolved main-sequence
stars.  A substantial number of binaries and NSs are also
ejected from the core during dynamical encounters.


\section{Summary}


We have carried out two complementary population synthesis studies of the
retention of neutron stars and the formation of neutron-star binaries in
globular clusters.  We have shown that with a ``conventional" distribution
of natal kick velocities, the fraction of neutron stars retained by a large
globular cluster can be as high as $8\%$ for neutron stars formed in
primordial binary systems.  However, this may be {\it in}sufficient to
explain the large numbers of neutron stars being discovered in, and
inferred for, clusters such as 47 Tuc.  A significant component of neutron
stars formed with substantially smaller (e.g., factors of $\sim2-3$) kick
speeds may be required in order to retain a sufficiently larger fraction of
the neutron stars (e.g., $\gtrsim 20\%$).

In a second population synthesis study, we have followed some $10^6$
primordial binaries and $10^4$ neutron stars through the history of a
globular cluster.  We find that such an initial population can be adequate
to account for the large measured and inferred population of close binary
millisecond pulsars in some of the massive, centrally concentrated
clusters.  With the current version of the population synthesis code we
also derive information on the numbers of NS-NS-, detached NS-WD-,
and WD-WD-binaries. Future studies, with an extended version of
the code, will allow us to compute in a more realistic way other branches
of the binary stellar evolution which we do not follow at the present time.



\end{document}